%% file: main.tex
\documentclass[9pt,conference]{IEEEtran}
\IEEEoverridecommandlockouts

\usepackage{cite}
\usepackage{amsmath,amssymb,amsfonts}
\usepackage{algorithmic}
\usepackage{graphicx}
\usepackage{textcomp}
\usepackage{xcolor}
\usepackage{flushend}
\usepackage{comment}
\usepackage{subcaption}
\usepackage{hyperref}
\input{helpers}

\def\BibTeX{{\rm B\kern-.05em{\sc i\kern-.025em b}\kern-.08em
    T\kern-.1667em\lower.7ex\hbox{E}\kern-.125emX}}
\definecolor{M5}{HTML}{1F77B4}
\definecolor{RAWNET2}{HTML}{AEC7E8}
\definecolor{RESNETSPEC}{HTML}{17BECF}

\begin{document}
\title{FakeMusicCaps: a Dataset for Detection and Attribution of Synthetic Music Generated via Text-to-Music Models \\
{\thanks{This material is based on research sponsored by the Defense Advanced Research Projects Agency (DARPA) and the Air Force Research Laboratory (AFRL) under agreement number FA8750-20-2-1004. The U.S. Government is authorized to reproduce and distribute reprints for Governmental purposes notwithstanding any copyright notation thereon. The views and conclusions contained herein are those of the authors and should not be interpreted as necessarily representing the official policies or endorsements, either expressed or implied, of DARPA and AFRL or the U.S. Government. This work is supported by the European Union under the Italian National Recovery and Resilience Plan (NRRP) of NextGenerationEU (PE00000001 - program ``RESTART'', PE00000014 - program ``SERICS''). This work is supported by the FOSTERER project, funded by the Italian Ministry of University, and Research within the PRIN 2022 program.}
}}

\author{\IEEEauthorblockN{Luca Comanducci}
\IEEEauthorblockA{
\textit{Politecnico di Milano}\\
Milan, Italy \\
luca.comanducci@polimi.it}
\and
\IEEEauthorblockN{Paolo Bestagini}
\IEEEauthorblockA{
\textit{Politecnico di Milano}\\
Milan, Italy \\
paolo.bestagini@polimi.it}
\and
\IEEEauthorblockN{Stefano Tubaro}
\IEEEauthorblockA{
\textit{Politecnico di Milano}\\
Milan, Italy \\
stefano.tubaro@polimi.it}
}

\author{\IEEEauthorblockN{Luca Comanducci, Paolo Bestagini, Stefano Tubaro} \IEEEauthorblockA{
Dipartimento di Elettronica, Informazione e Bioingegneria (DEIB), Politecnico di Milano\\
Piazza Leonardo Da Vinci 32, 20133 Milan, Italy\\
Email: name.surname@polimi.it
}
}

\maketitle

\begin{abstract}
Text-To-Music (TTM) models have recently revolutionized the automatic music generation research field. Specifically, by reaching superior performances to all previous state-of-the-art models and by lowering the technical proficiency needed to use them. Due to these reasons, they have readily started to be adopted for commercial uses and music production practices. This widespread diffusion of TTMs poses several concerns regarding copyright violation and rightful attribution, posing the need of serious consideration of them by the audio forensics community. In this paper, we tackle the problem of detection and attribution of TTM-generated data. We propose a dataset, FakeMusicCaps that contains several versions of the music-caption pairs dataset MusicCaps re-generated via several state-of-the-art TTM techniques. We evaluate the proposed dataset by performing initial experiments regarding the detection and attribution of TTM-generated audio.
\end{abstract}

\begin{IEEEkeywords}
Music Generation, Text-To-Music, Audio Forensics, DeepFake
\end{IEEEkeywords}

\section{Introduction}
\label{sec:intro}

Deep learning-based music generation~\cite{briot2020deep} has been recently revolutionized by the introduction of Text-To-Music models. 
TTM models are usually either based on a language model that decodes continuous or discrete tokenized embeddings obtained via some neural audio codec~\cite{kumar2024high,defossez2023high}, such as MusicLM~\cite{agostinelli2023musiclm}, MusicGEN~\cite{copet2024simple}, MAGNeT~\cite{ziv2024masked} and JASCO~\cite{tal2024joint} or on latent diffusion models operating on some compressed form of audio, such as AudioLDM~\cite{liu2023audioldm}, AudioLDM2~\cite{liu2024audioldm}, MusicLDM~\cite{chen2024musicldm}, Noise2Music~\cite{huang2023noise2music}, Mustango~\cite{melechovsky2024mustango}.

These models are characterized by being good in terms of performance and also simple to use, lowering the technical proficiency needed to successfully interact with them~\cite{ronchini2024paguri}. This combination of factors has made them extremely appetible to the general public and of interest by private industries. 

In Fact several commercial TTMs have been proposed, such as Suno~\cite{sunoSuno} (reaching the record for the biggest investment ever in an AI music startup, namely \$125 million) and Udio~\cite{udioUdioMusic}. Recently, both these companies have been sued by major record companies and have consecutively admitted to copyright infringement, by training their respective models also using unlicensed music. As both the capabilities and the commercial interest of these models grow it is becoming increasingly necessary to try to develop forensic approaches to detect and analyze music generated via TTMs~\cite{feffer2023deepdrake}.

The multimedia forensics field is well mature in fields related to image~\cite{sha2023fake,ning2019Attributing, corvi2023detection, abady2024siamese, wissmann2024whodunit} and video~\cite{mandelli2019facing} deepfake detection and model attribution, but in the audio domain has been almost exclusively applied to speech signals~\cite{wu2024codecfake, salvi2022exploring,  bhagtani2023synthesized}.

In the music domain, most efforts have concentrated on Singing voice detection~\cite{zang2024singfake,xie2024fsd,chen2024singing} with the development of specific challenges~\cite{guragain2024speechfoundationmodelensembles} to tackle the problem. An approach to music deepfake detection has been proposed in~\cite{afchar2024detecting}, where however no fake music is considered, instead the authors focus on detecting with which neural audio codec real music tracks are processed.

Research in this field is also limited by economic reasons since most models are developed by tech giants that often do not release the code and/or weights. Also, available paired text-music dataset except for MusiCaps~\cite{agostinelli2023musiclm}, Song Describer~\cite{manco2023song} and MusicBench~\cite{melechovsky2024mustango}.

In this paper, we propose the \textit{FakeMusicCaps dataset}, with the objective of encouraging research in music deepfake detection. To build FakeMusicCaps, we replicate MusicCaps by using its captions as input to Five state-of-the-art TTM models, namely MusicGen~\cite{copet2024simple}, MusicLDM~\cite{chen2024musicldm}, AudioLDM2~\cite{liu2024audioldm}, Stable Audio Open~\cite{evans2024stable} and Mustango~\cite{melechovsky2024mustango}. The nature of FakeMusicCaps makes it easy to incorporate future TTMs by simply generating music examples using the same procedure. We perform a benchmark study, on FakeMusicCaps, by studying if it is possible to perform detection and attribution, i.e. classifying the input music as either real or belonging to one of the chosen TTM models. We analyze how the model performs both in closed set or open set scenarios, where the latter also includes data belonging to generators not seen during training, which in this paper belong to the SunoCaps dataset~\cite{civit2024sunocaps}.

At the same time of this work, a similar dataset, named SONICS, has been proposed~\cite{rahman2024sonics}, which however considers only the commercially-available models Suno and Udio and performs only real/fake music detection. We instead focus on open-source TTMs and consider commercial ones only in the open set classification. The reasoning behind this is that open-source techniques are possibly available to a wider part of the research community, which could use them to integrate FakeMusicCaps as they see fit.

The code used to generate FakeMusicaps and perform the experiments\footnote{\url{https://github.com/polimi-ispl/FakeMusicCaps}} as well as the full dataset\footnote{\url{https://zenodo.org/records/13732524}} are publicly available.
The rest of the paper is organized as follows. In Sec.~\ref{sec:probform} we introduce the attribution problem for TTM models. In Sec.~\ref{sec:dataset} we describe how the FakeMusicCaps dataset was created. Sec.~\ref{sec:expanalysis} presents the experimental setup used to conduct the experiments, while in Sec.~\ref{sec:expresults} we present the results aimed analyzing the complexity of TTM attribution and the effectiveness of the dataset. Finally, in Sec.~\ref{sec:conclusion} we draw some conclusions.

\section{Problem Formulation}
\label{sec:probform}
\begin{figure}[htb]
    \centering
    \includegraphics[width=\linewidth]{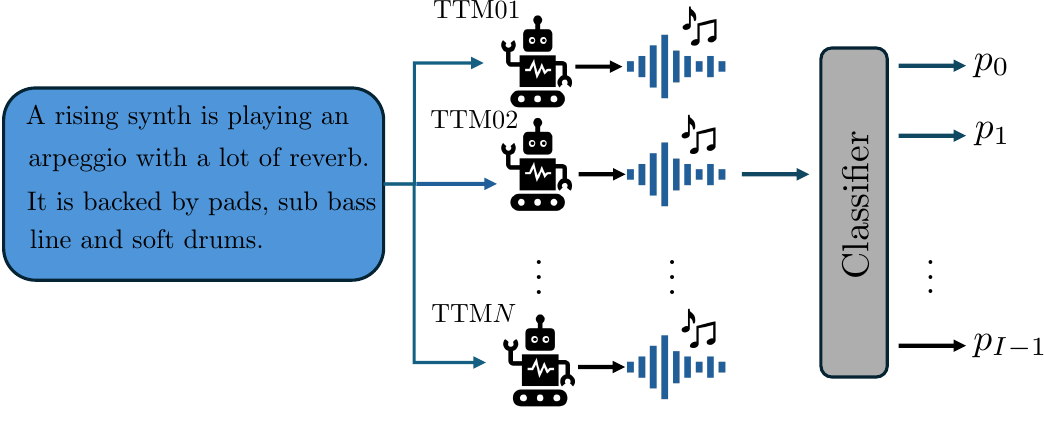}
    \caption{Schematic representation of the text-to-music attribution problem.}
    \label{fig:overview-scheme}
\end{figure}

Given some kind of text representation $\tau$ and a composite model $\mathcal{T}(\cdot)$, TTMs techniques model the function $\mathbf{x} =\mathcal{T}(\tau)$, where $\mathbf{x}\in\mathbb{R}^{1 \times N}$ is an audio waveform containing music corresponding to the textual description provided in $\tau$.
Then, the text-to-music attribution problem, schematically shown in Fig.~\ref{fig:overview-scheme}, can be formally defined as follows. Given the time-discrete music signal $\mathbf{x}$ and a set of $I$ TTM models $\{ \mathcal{T}_0, \ldots, \mathcal{T}_{I-1}\}$, the objective is to retrieve which generator $\mathcal{T}_i$ has been used to generate $\mathbf{x}$. This is done by training a classifier that takes as input $\mathbf{x}$ and outputs the probabilities $p_i, i=0,\ldots, I-1$ of $\mathbf{x}$ being generated using each of the known TTM models.

The attribution problem is often considered in both closed- and open-set scenarios. In the former, all generators are seen both during training and testing, while in the latter, some TTMs are \textit{unknown} during training and seen only at testing time, posing the need to develop specific classification strategies.

\section{FakeMusicCaps Dataset}
\label{sec:dataset}
In this section, we describe how the FakeMusicCaps dataset was created, first by presenting the chosen TTM models and then describing the generation procedure.

\subsection{Considered Architectures}
In this Section, we present an overview of the architectures (TTM01-TTM05) used to create the FakeMusicCaps dataset.

\textbf{TTM01-MusicGen}~\cite{copet2024simple} is an autoregressive language model, based on a single-stage transformer that decodes discrete audio tokens obtained via Encodec~\cite{defossez2023high}. It was trained over an undisclosed dataset of over 20K hours of music.  We use the \textit{medium} checkpoint consisting of $1.5B$ parameters.

\textbf{TTM02-MusicLDM}~\cite{chen2024musicldm} is a latent diffusion model operating on compressed audio representations extracted via HiFi-GAN~\cite{kong2020hifi}. It adapts AudioLDM to the musical domain, by introducing beat-synchronous audio mixup and beat-synchronous latent mixup strategies to augment the quantity of data used for training. 
The text conditioning is provided via CLAP~\cite{wu2023large}, which the authors finetune on music for a total of 20K hours. The MusicLDM model is then trained on the Audiostock dataset~\cite{wu2023large}, containing 455.6 hours of music.

\textbf{TTM03-AudioLDM2}~\cite{liu2024audioldm} 
is a latent diffusion model where the audio is compressed via a Variational AutoEncoder (VAE) and HiFiGAN, similarly to AudioLDM. However, the major difference of AudioLDM2, is that the diffusion model is conditioned through AudioMAE~\cite{huang2022masked} that enables the adoption of a ``Language of Audio'', that enables to generate a wide variety of typoes of audio.  We use the \textit{audioldm2-music} checkpoint to build FakeMusicCaps, specifically trained for text-to-music generation.

\textbf{TTM04-Stable Audio Open}~\cite{evans2024stable} is a latent-diffusion architecture generating stereo data at $44.1~\mathrm{kHz}$ based on a variant of Stable Audio~\cite{evans2024long} that uses T5~\cite{raffel2020exploring} as a text encoder. The model is trained only on Creative Commons-licensed audio data for a total of 7.3K hours of audio.

\textbf{TTM05-Mustango}~\cite{melechovsky2024mustango} is a diffusion-based TTM model that through a Music-domain-knowledge-informed UNet (MuNet) injects music concepts such as chord, beats, key or tempo in the generated music, during the reverse diffusion process. Through data augmentation, the authors generate the MusicBench dataset, composed of 53.168 tracks, to train the model. The model generates at $16~\mathrm{kHz}$

\subsection{Generation strategy}
In this section we describe the strategy used to generate the FakeMusicCaps dataset. We consider the MusicCaps~\cite{agostinelli2023musiclm} dataset, consisting of 5.5k 10 seconds of music clips from AudioSet~\cite{gemmeke2017audio}, each one paired with an annotation by a professional musician. MusicCaps has rapidly become the benchmark dataset for the evaluation of TTM models. To create FakeMusicCaps, we use the caption from MusicCaps and for each one of them we generate a corresponding 10-second audio track using models (TTM01-TTM05) for a total of 27605 music tracks corresponding to almost 77 hours. Since the objective of the dataset is to provide an initial dataset for the analysis of the detection and attribution of music generated via TTM models, we adopt an audio pipeline that ensures that all audios are represented using the same format. Specifically, each track is converted to mono and downsampled to $F_s=16~\mathrm{kHz}$. Finally,  we save each track using the 32-bit float wav format.

\section{Experimental Analysis}
\label{sec:expanalysis}
In this section ,we describe the techniques and setup used during the experiment aimed at showing a first validation of the FakeMusicCaps dataset.

\subsection{Dataset}
We used the FakeMusicCaps dataset during the training and test procedures. In order to build a test set disjointed from the training one, we selected $320$ tracks from FakeMusicCaps, selecting, for each TTM model, the ones having the same captions of the SunoCaps~\cite{civit2024sunocaps} dataset. This choice was operated in order to be able to coherently use the Suno-generated music excerpts from SunoCaps to perform the open-set scenario experiments. 

\subsection{Baselines}
We use three classification models as simple benchmarks of the FakeMusicCaps for deepfake music detection and attribution. 
We first considered a very simple network operating on raw audio, namely M5~\cite{dai2017very}. This network consists of only 0.5M parameters and is used as an initial experiment to see if the problem is even worth of attention.

Then, we selected a more complicated method operating on raw audio, RawNet2~\cite{tak2021end} and on log-spectrograms, denoted ResNet18+Spec~\cite{he2016deep}.
RawNet2 is an end-to-end neural network that has been used as a baseline for several antispoofing challenges such as ASVspoof 2021 and consists of Sinc Filters, followed by residual blocks and a Gated Recurrent Unit (GRU). 

Finally, we made use of ResNet18. This is an 18-layer deep convolutional layer with residual connections, we modify it making it suitable to work with 1-channel log-spectrograms.

All methods were modified by adding a fully connected layer with a number of neurons at the end of the networks, corresponding to the number of considered classes.

\subsection{Training}
All models were trained to discriminate between $6$ different classes, comprising the $5$ known TTM models and the real music signals belonging to MusicCaps.
We trained all models using the cross-entropy as a loss function and used the Adam optimizer with a learning rate of $1e-4$.
All networks were traineto set a maximum of $100$ epochs, ending the training earlier if the loss did not improve for more than $10$ consecutive epochs. We used batch size of $32$ for M5 and $16$ for both RawNet2 and ResNet18 + Spec. In the case of ResNet18 + Spec, we computed the STFT with 512 frequency points, using a hann window of length 512 samples with hop size 128 samples.

\subsection{Classification Techniques}
In the \textit{closed set} classification problem, given a raw audio waveform corresponding to music, we want to identify the generation method from the set of \textit{known} (i.e. a set of models included in the training dataset) TTM models, while in the case of the \textit{open set} cclassification we want to detect it from a TTM model that it is \textit{unknown}, i.e. not included in the training dataset.
If we consider $p_i$ as the output of the softmax layer of the models, then in the closed set case, class attribution can simply be performed by computing $\arg \max_i p_i$. 
For open set classification, instead, we follow two different approaches. In the open set \textit{(threshold)} technique~\cite{hendrycks2022baseline} we compute the two highest values of $p_i$ and then classify the input example as unknown if the ratio between these values is higher than a threshold. The rationale is that if the method was known, only one $p_i$ value should be high. We choose a threshold of $2$ following~\cite{salvi2022exploring}.
In the open set \textit{SVM} technique, instead, we train a one-class Support Vector Machine (SVM) using radial basis functions kernel over the $p_i$ values computed over the training data. The output of the classification is binary: either the class is known or not.

\section{Results}
\label{sec:expresults}
In this section, we present preliminary results aimed at demonstrating the suitability of FakeMusicCaps as an initial dataset for text-to-music model detection and attribution.

\subsection{Closed-Set Performances}
Despite closed-set classification on a single dataset is often considered a trivial task in forensic applications, it is worth investigating the performance of the tested methods in this scenario.
Table~\ref{tab:closed_set} reports closed-set classification results in terms of balanced accuracy $\mathrm{ACC}_B$, Precision, Recall, and F1 Score. The left column of Fig.~\ref{fig:cm_visualization} shows the confusion matrix corresponding to M5, RawNet2 and ResNet18+Spec, respectively. In all metrics, the best performances are obtained by ResNet18 + Spec, while RawNet2 obtains slightly worse results than M5. From the inspection of the confusion matrices, we can see that ResNet18 slightly confounds TTM03 (AudioLDM2) with TTM05(Mustango), both diffusion-based models. M5 has slightly lower performances in detecting the ground-truth data, while RawNet2 struggles more to detect model TTM02 (MusicLDM). 
\begin{table}[htb]
\centering
    \caption{Closed-set classification performances.}
    \resizebox{\columnwidth}{!}{\begin{tabular}{lcccc}
        \hline
        Model & $\mathrm{ACC_B} \downarrow$ &  Precision & Recall &F1 Score \\
        \hline
        \hline 
M5 &0.90 &0.90&0.90& 0.90\\
RawNet2 &0.88 &0.89&0.88& 0.88\\
ResNet18 + Spec & $\mathbf{1.00}$ &$\mathbf{1.00}$&$\mathbf{1.00}$& $\mathbf{1.00}$\\
\hline
\end{tabular}}
\label{tab:closed_set}
\end{table}

\subsection{Open Set Performances}
\begin{figure*}[t]
\centering
\begin{minipage}[c]{0.3\linewidth} 
  \centering
\centerline{\includegraphics[width=\columnwidth]{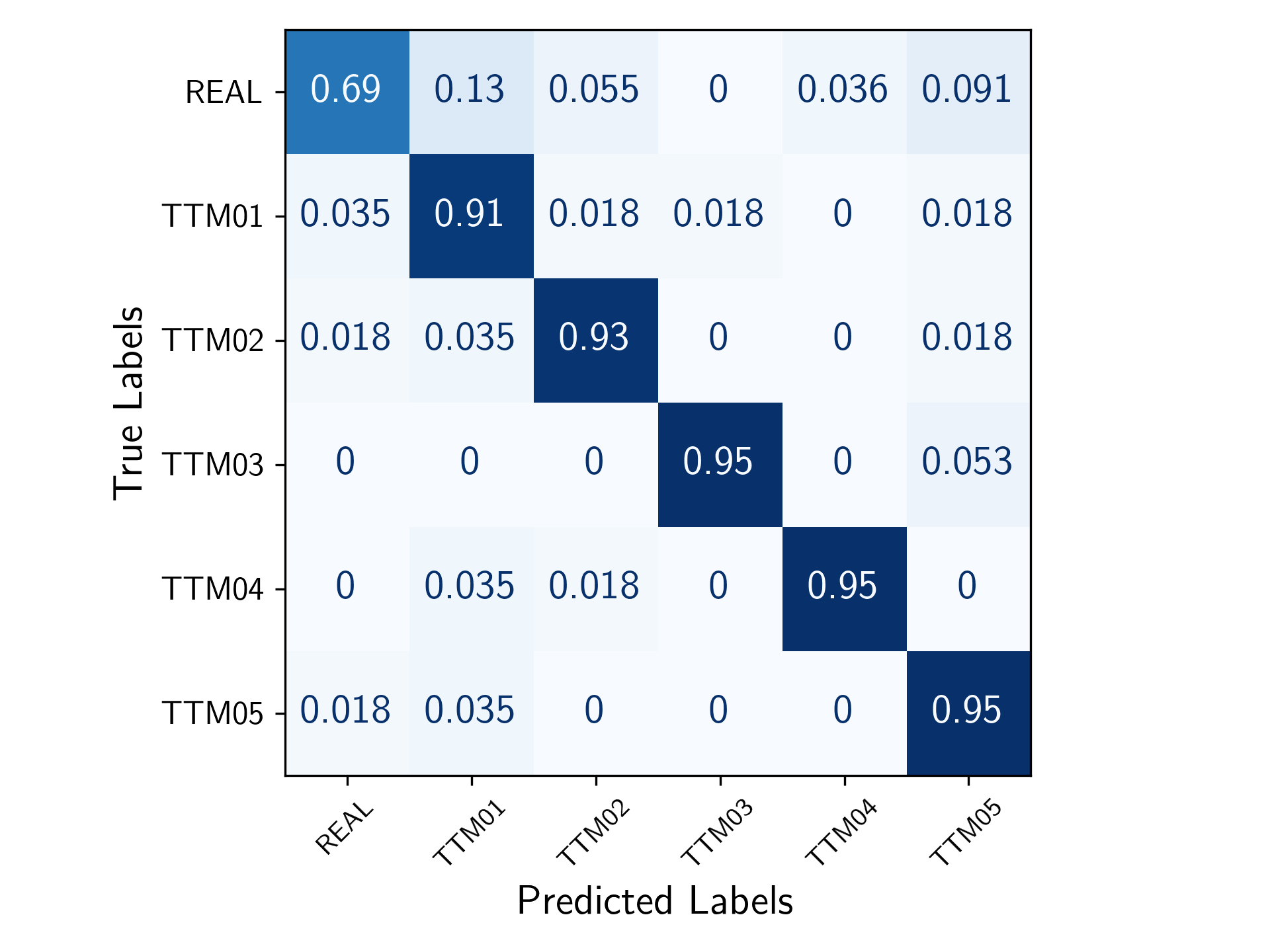}}
\end{minipage}
\begin{minipage}[c]{0.3\linewidth}
  \centering
\centerline{\includegraphics[width=\columnwidth]{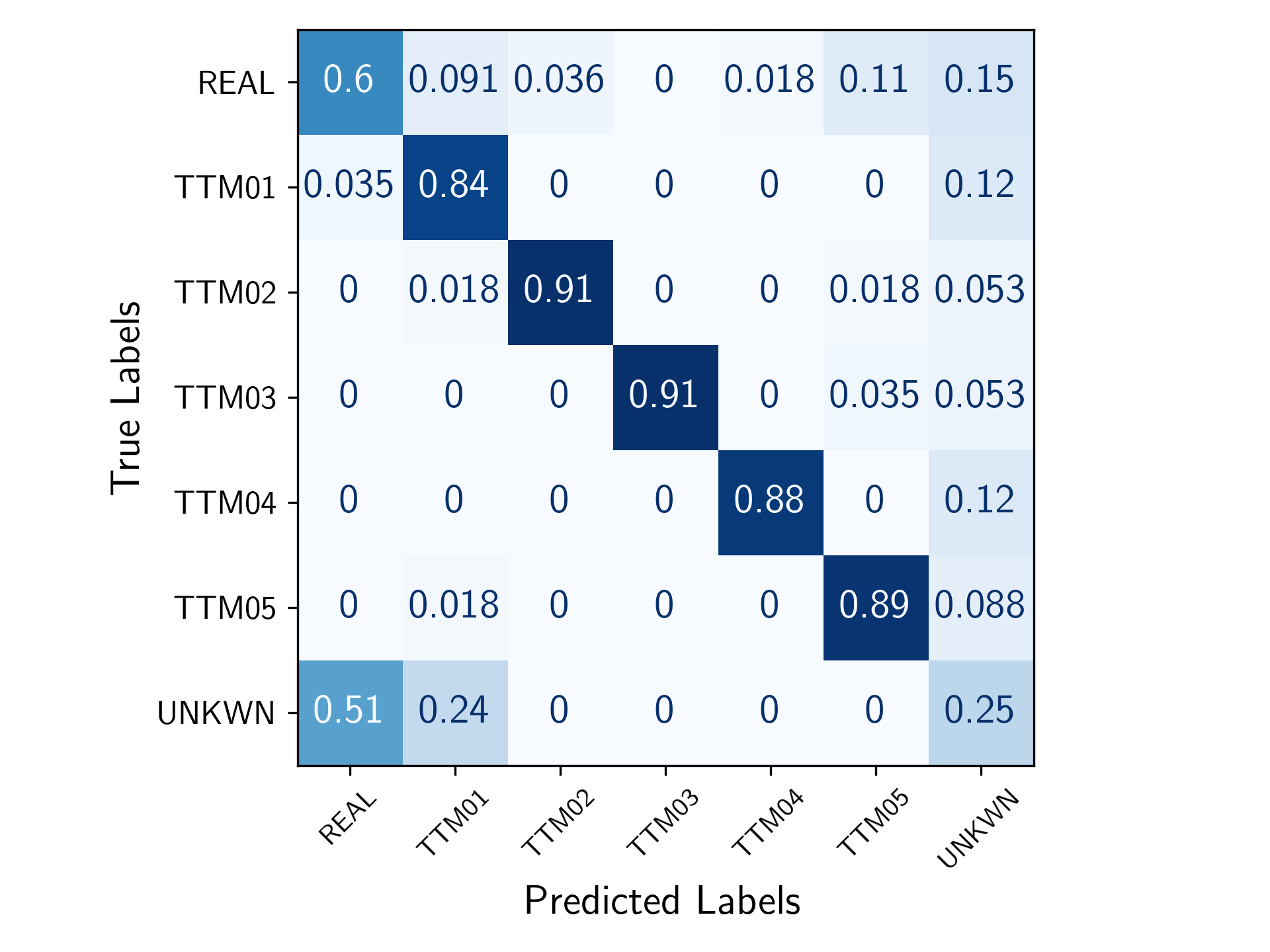}}
\end{minipage}
\begin{minipage}[c]{0.3\linewidth}
  \centering
\centerline{\includegraphics[width=\columnwidth]{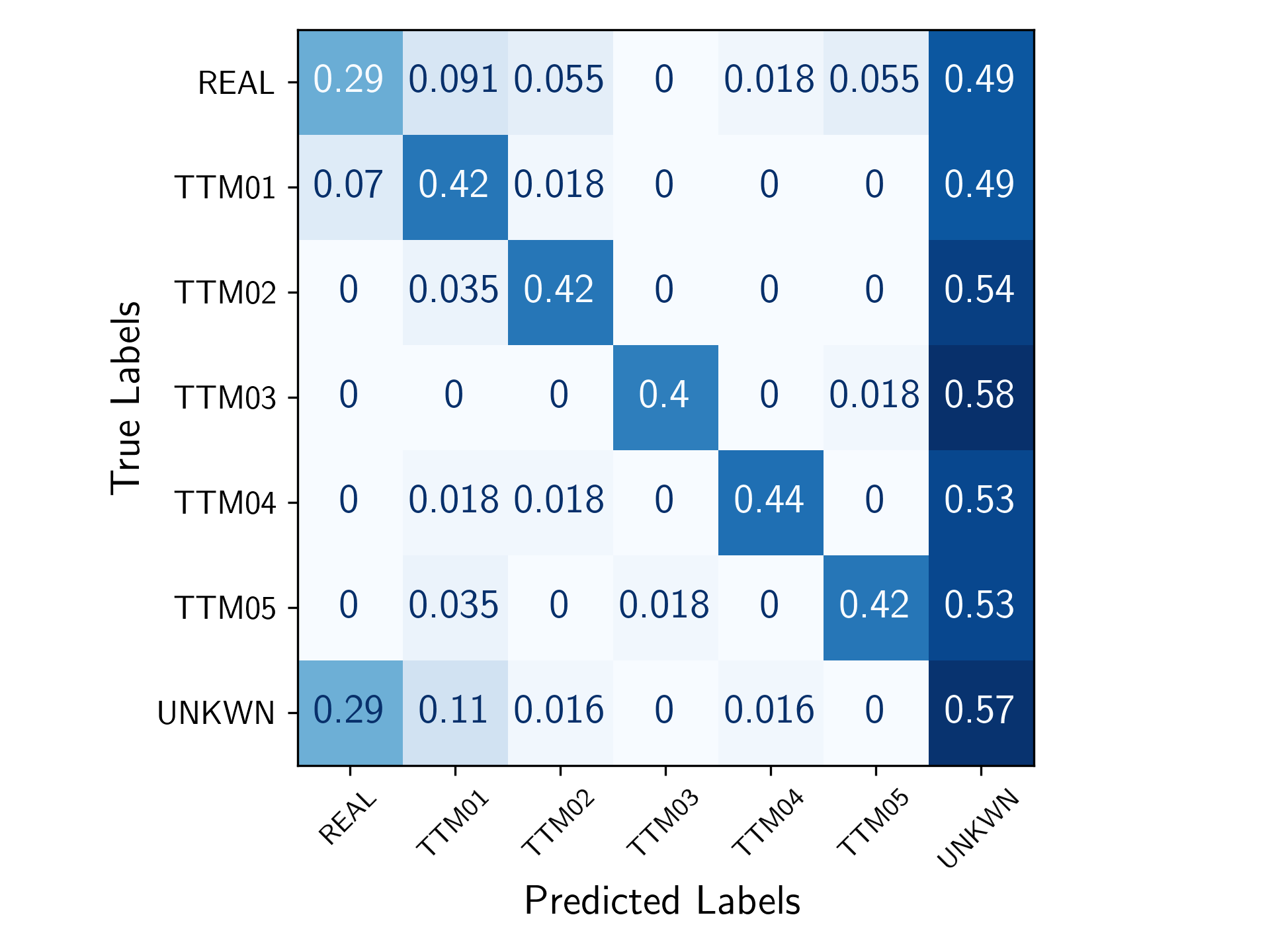}}
\end{minipage}
\vfill
\begin{minipage}[c]{0.3\linewidth}
  \centering
\centerline{\includegraphics[width=\columnwidth]{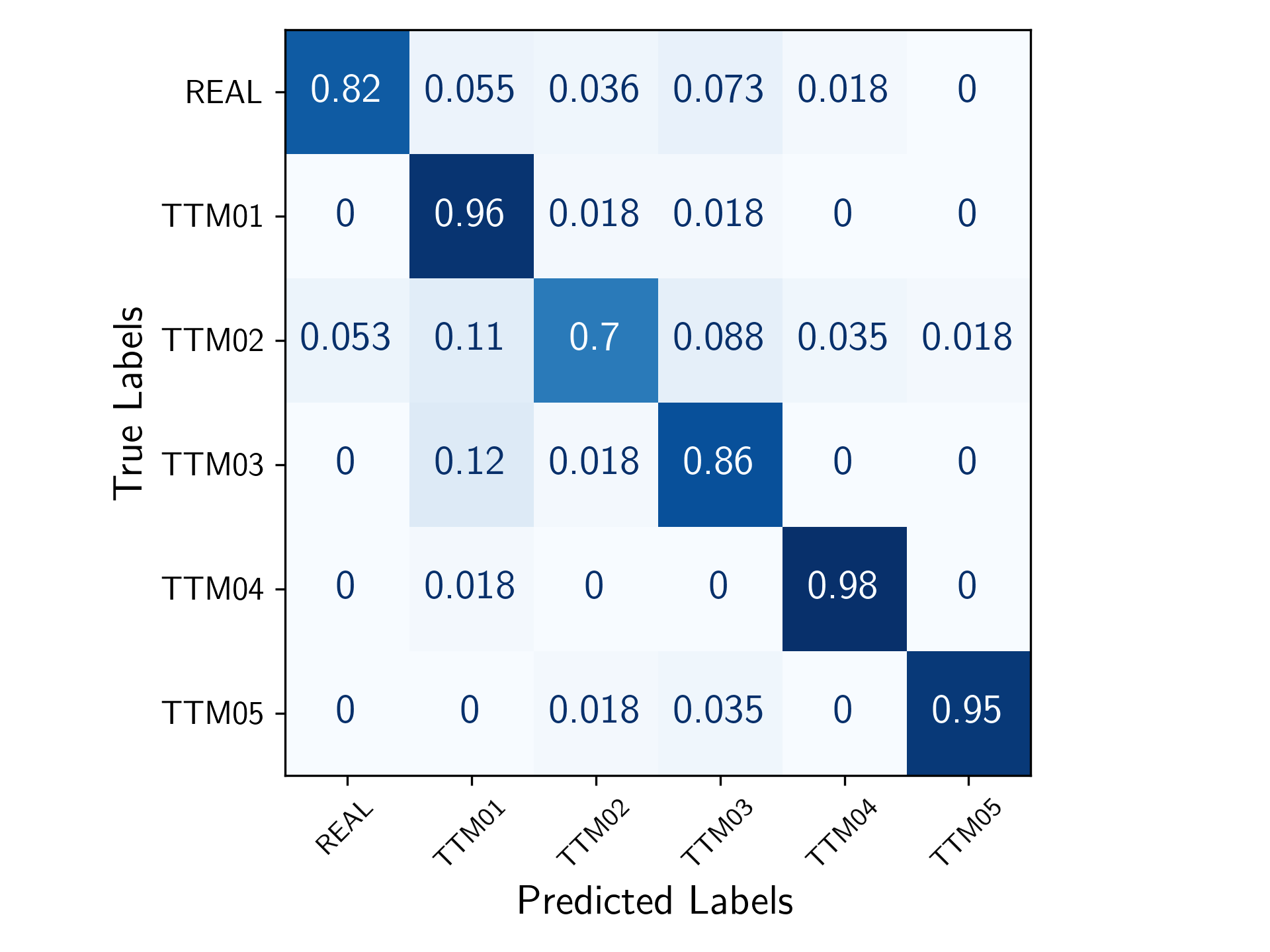}}
\end{minipage}
\begin{minipage}[c]{0.3\linewidth}
  \centering
\centerline{\includegraphics[width=\columnwidth]{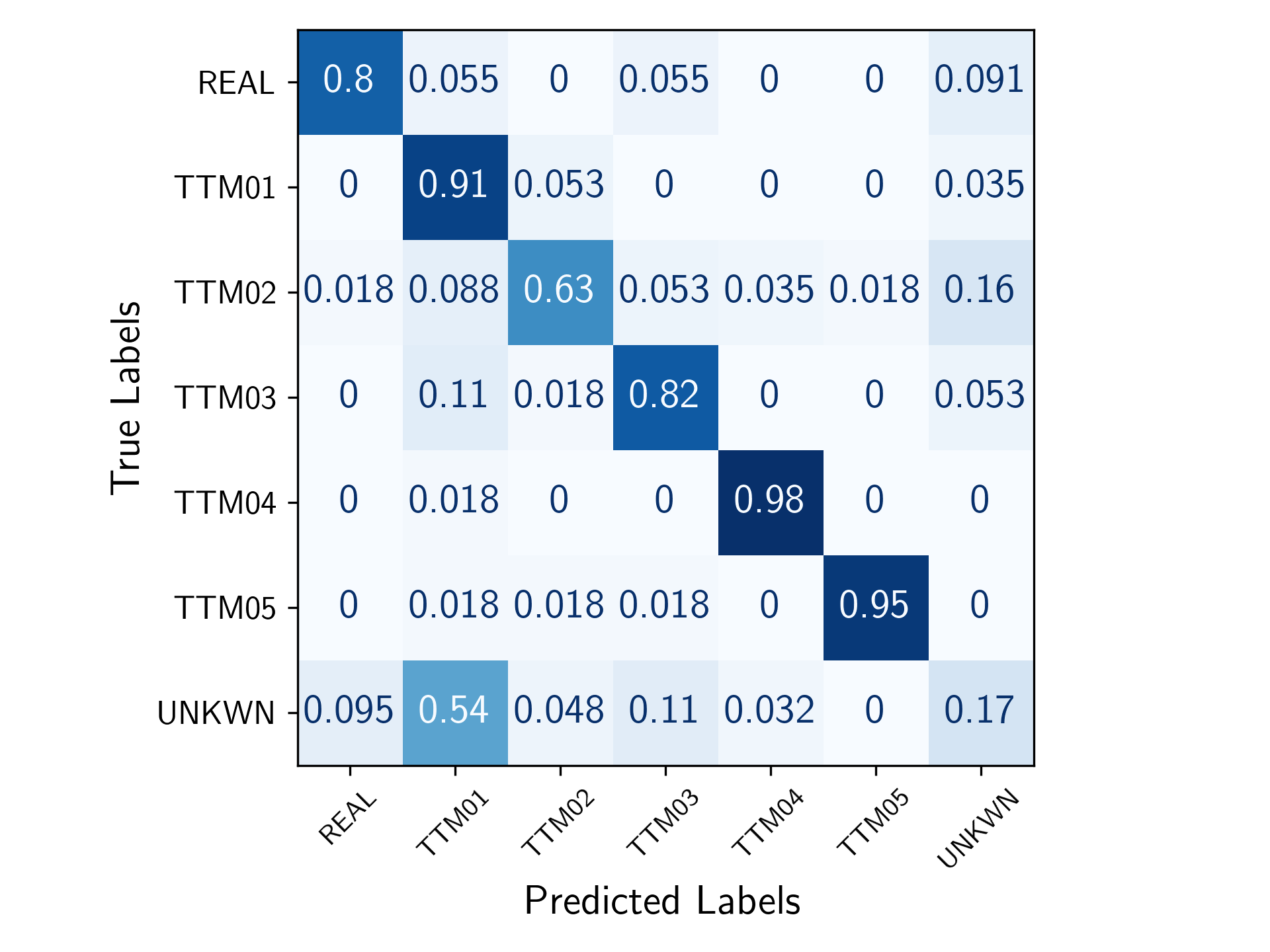}}
\end{minipage}
\begin{minipage}[c]{0.3\linewidth}
  \centering
\centerline{\includegraphics[width=\columnwidth]{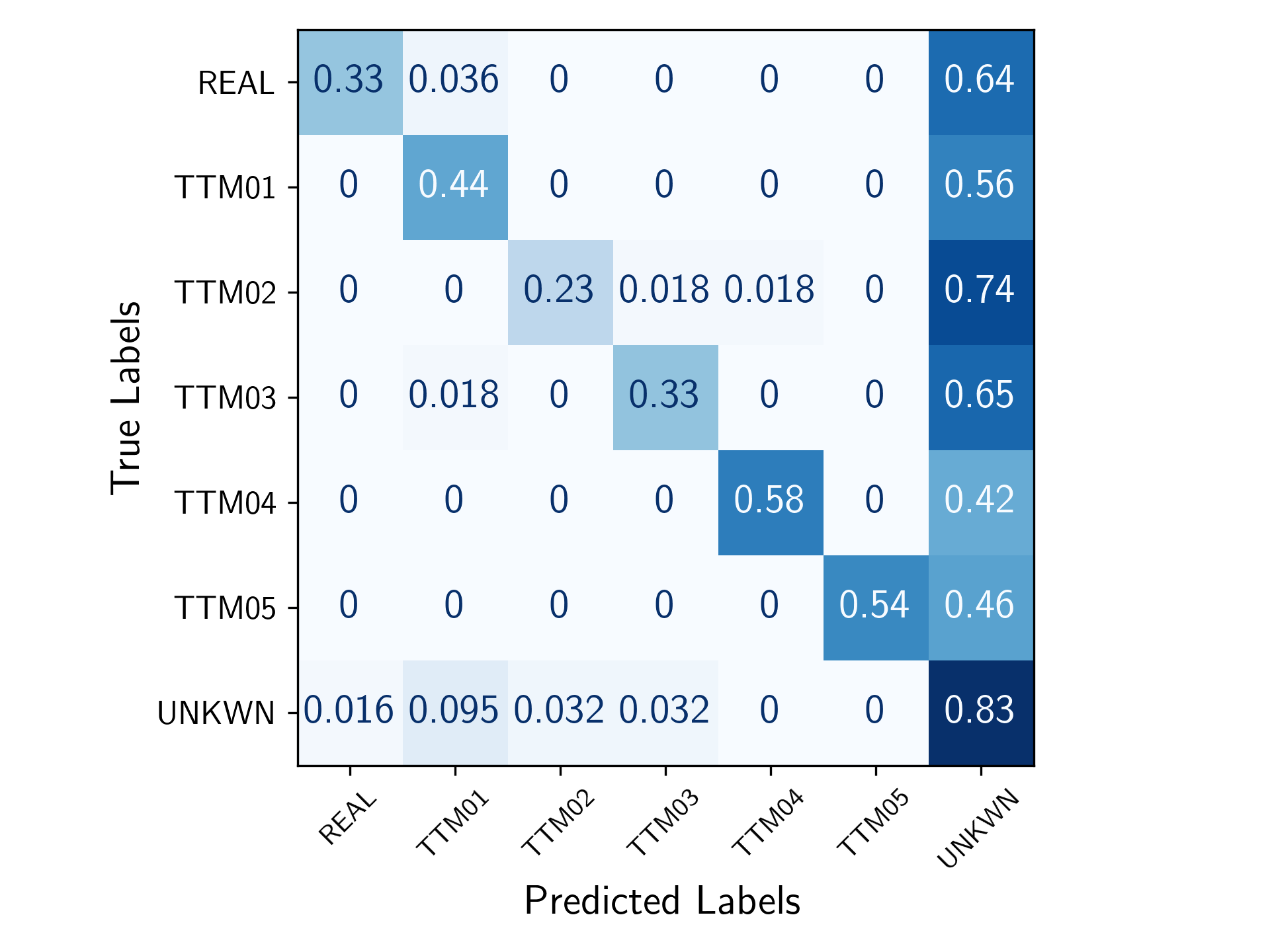}}
\end{minipage}
\vfill
\begin{minipage}[c]{0.3\linewidth}
  \centering
\centerline{\includegraphics[width=\columnwidth]{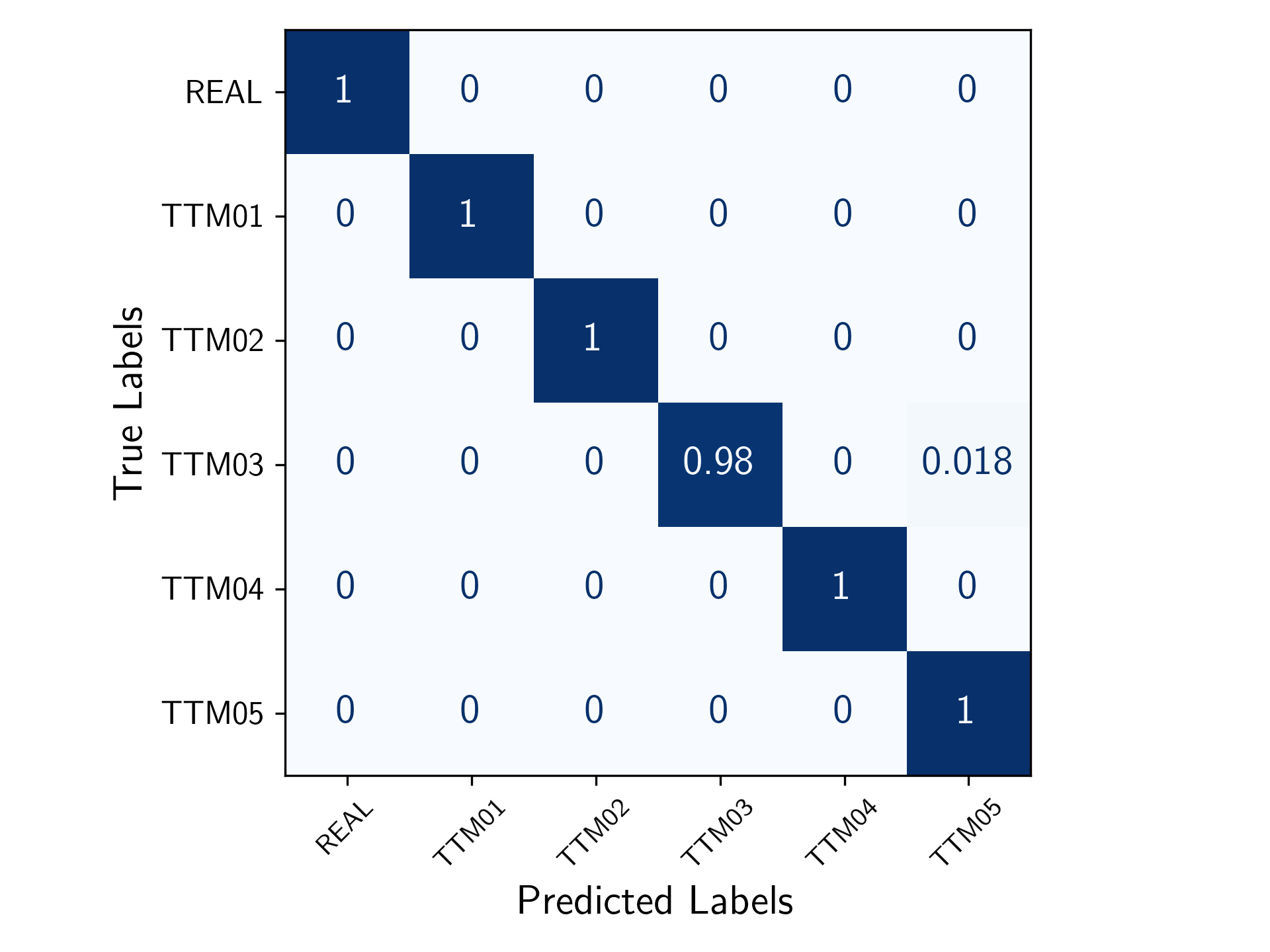}}
  \centerline{(a) Closed set}
\end{minipage}
\begin{minipage}[c]{0.3\linewidth}
  \centering
\centerline{\includegraphics[width=\columnwidth]{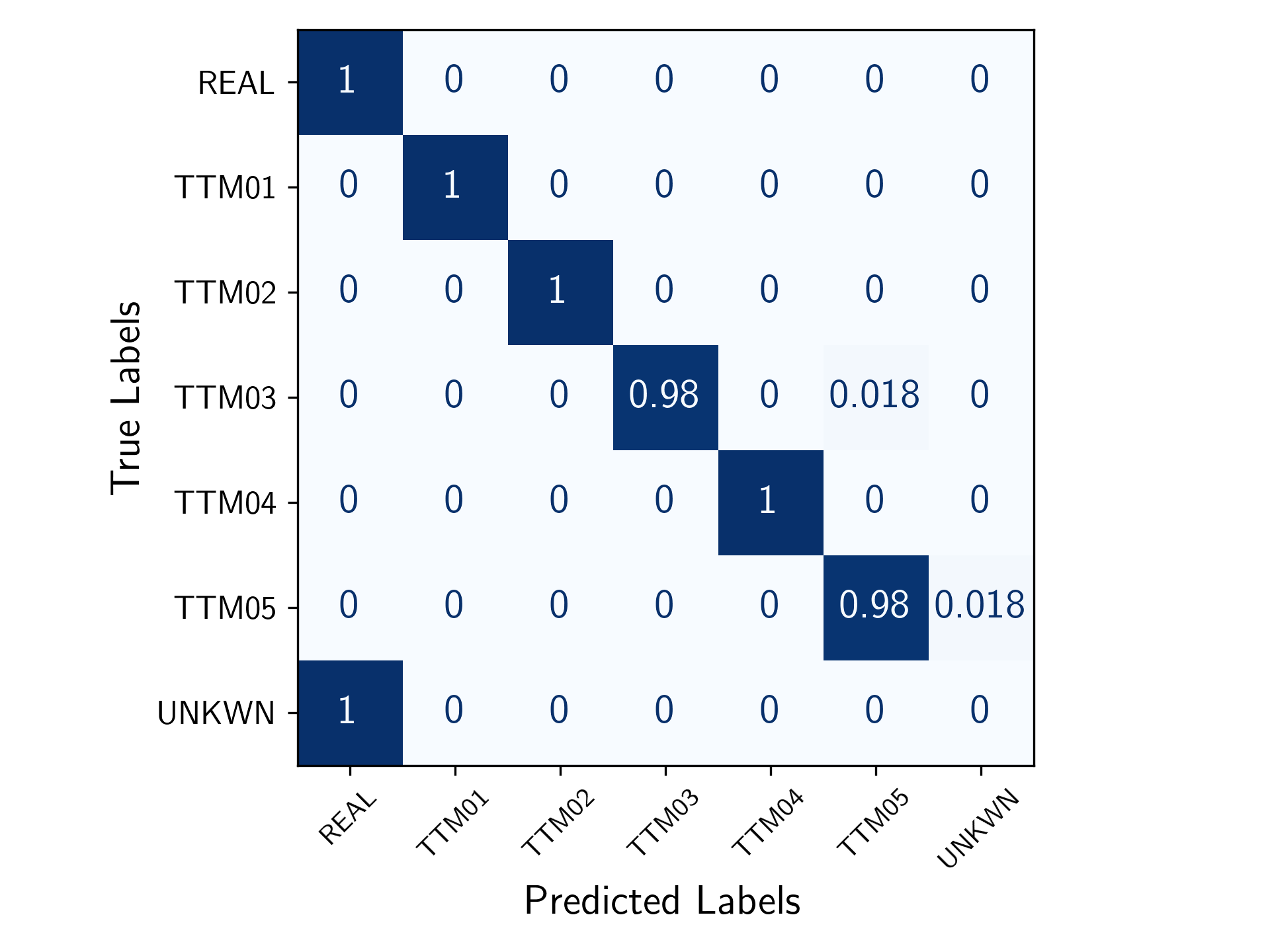}}
  \centerline{(b) Open set - threshold}
\end{minipage}
\begin{minipage}[c]{0.3\linewidth}
  \centering
\centerline{\includegraphics[width=\columnwidth]{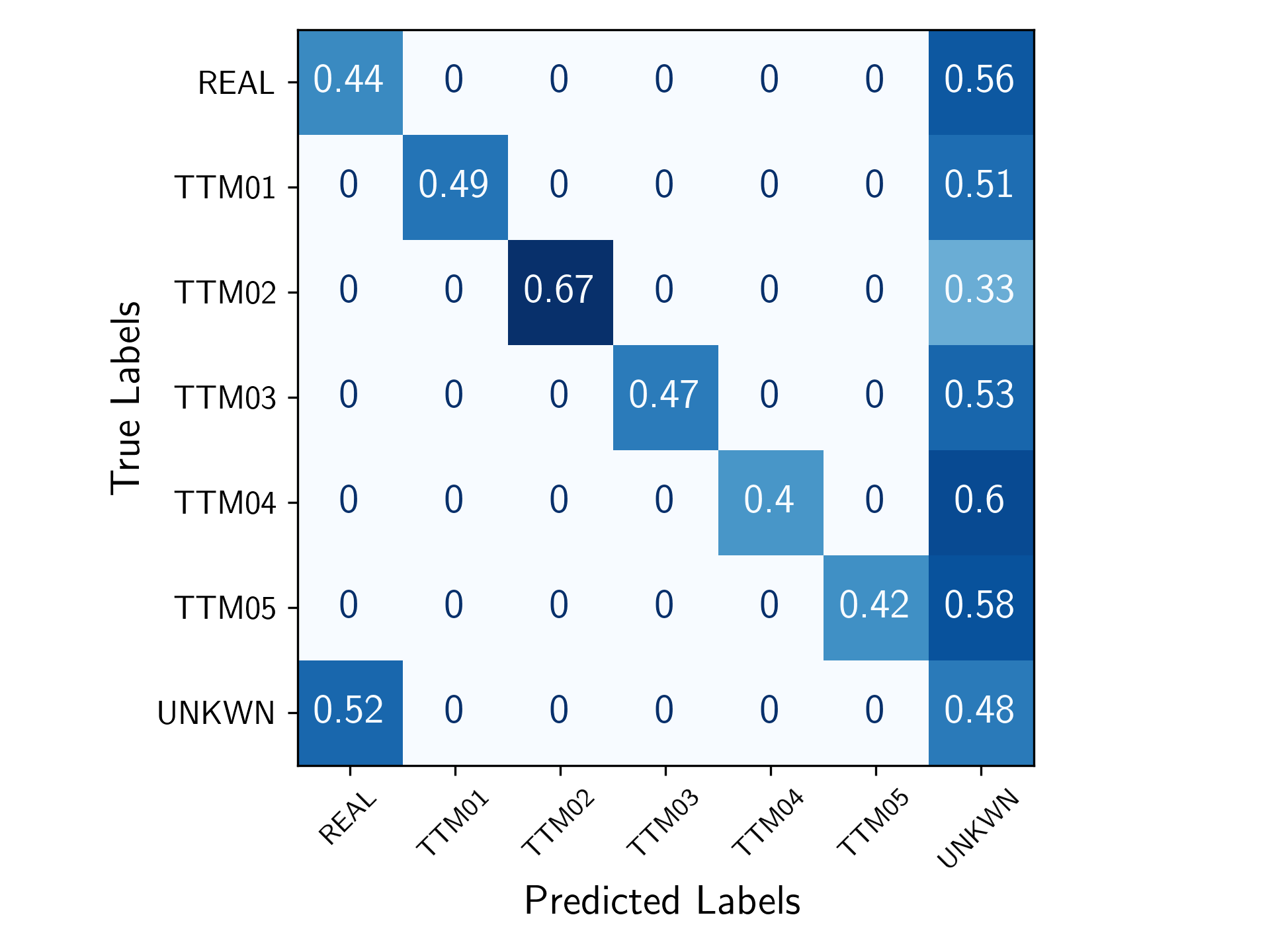}}
  \centerline{(c) Open set - SVM}
\end{minipage}
\caption{Confusion matrices of M5 (top), RawNet2 (middle) and ResNet+Spec (bottom) in the three classification scenarios.}
\label{fig:cm_visualization}
\end{figure*}

We show in Table~\ref{tab:open_set_thresh} the performances when performing open set classification using the threshold approach and the corresponding confusion matrices in the second column of Fig.~\ref{fig:cm_visualization}.
\begin{table}[htb]
\centering
    \caption{Open set (Threshold) classification performances.}
    \resizebox{\columnwidth}{!}{\begin{tabular}{lcccc}
        \hline
        Model & $\mathrm{ACC_B}$ &  Precision & Recall &F1 Score \\
        \hline
        \hline 
    M5 &0.76 &0.76&0.76& 0.75\\
    RawNet2 &0.75 &0.75&0.75& 0.74\\
    ResNet18 + Spec  &\textbf{0.85} &\textbf{0.78}&\textbf{0.85}& \textbf{0.8}\\
        \hline
        \end{tabular}}
\label{tab:open_set_thresh}
\end{table}
Results corresponding to the open set classification using the SVM approach are shown in Table~\ref{tab:open_set_svm}, with the corresponding confusion matrices in the last column of Fig.~\ref{fig:cm_visualization}. 
\begin{table}[htb]
\centering
    \caption{Open set (SVM) classification performances.}
    \resizebox{\columnwidth}{!}{\begin{tabular}{lcccc}
        \hline
        Model & $\mathrm{ACC_B}$ &  Precision & Recall &F1 Score \\
        \hline
        \hline 
M5 &0.42 &0.67&0.42& 0.48\\
RawNet2 &0.47 &0.80&0.47& 0.52\\
ResNet18 + Spec  &\textbf{0.48} &\textbf{0.80}&\textbf{0.48}& \textbf{0.56}\\
        \hline
        \end{tabular}}
\label{tab:open_set_svm}
\end{table}
As expected, in this case performance is much worse for all considered models. If we look at the results reported in the two tables, we can see that again ResNet18 + Spec obtains the best performance using both methods and that results obtained via the SVM technique are much worse than the ones obtained via the thresholding approach. The perspective however becomes different if we look at the confusion matrices. In the thresholding method, we can see that ResNet18 + Spec obtains the  best performance when classifying the known methods, but misclassifies all the audio excerpts belonging to the UNKNWN class. Interestingly enough, these are confounded with the real examples, which is somewhat expected, given that the commercial model Suno is probably the most realistic of the considered TTM models. M5 and RawNet2 obtain somehow similar performance, with the former confounding UNKNWN examples with real and MusicGen-generated ones, while the latter confounds them mostly with MusicGen.
In the case of the SVM approach all models have a different behavior, specifically approximately half the time, they mistake the known techniques for the unknown one. Interestingly, RawNet2 obtains the highest accuracy of $0.82$ for what concerns the unknown class, and even in this case ResNet18 mistakes it for the real one.

\subsection{Impact of window size}
\begin{figure*}[t]
\centering
\begin{minipage}[c]{0.32\linewidth}
  \centering
\centerline{\includegraphics[width=\columnwidth]{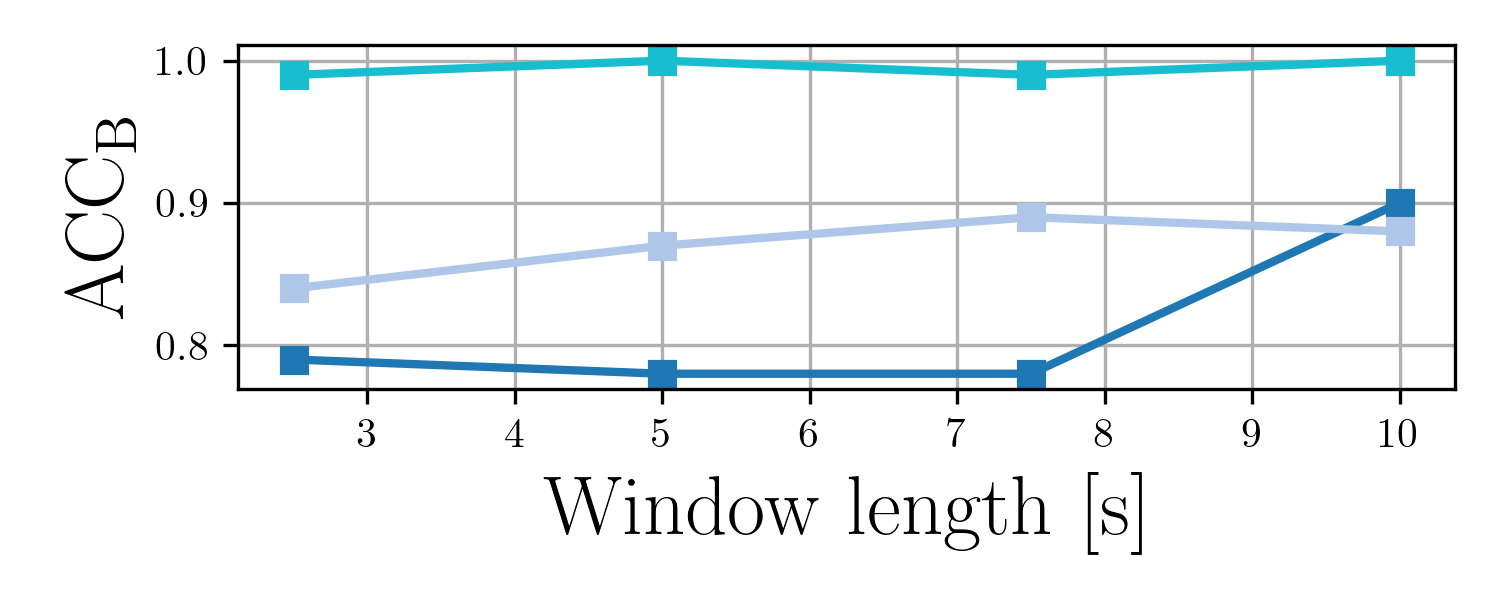}}
\centerline{(a) Closed set}
\end{minipage}
\hfill
\begin{minipage}[c]{0.32\linewidth}
  \centering
\centerline{\includegraphics[width=\columnwidth]{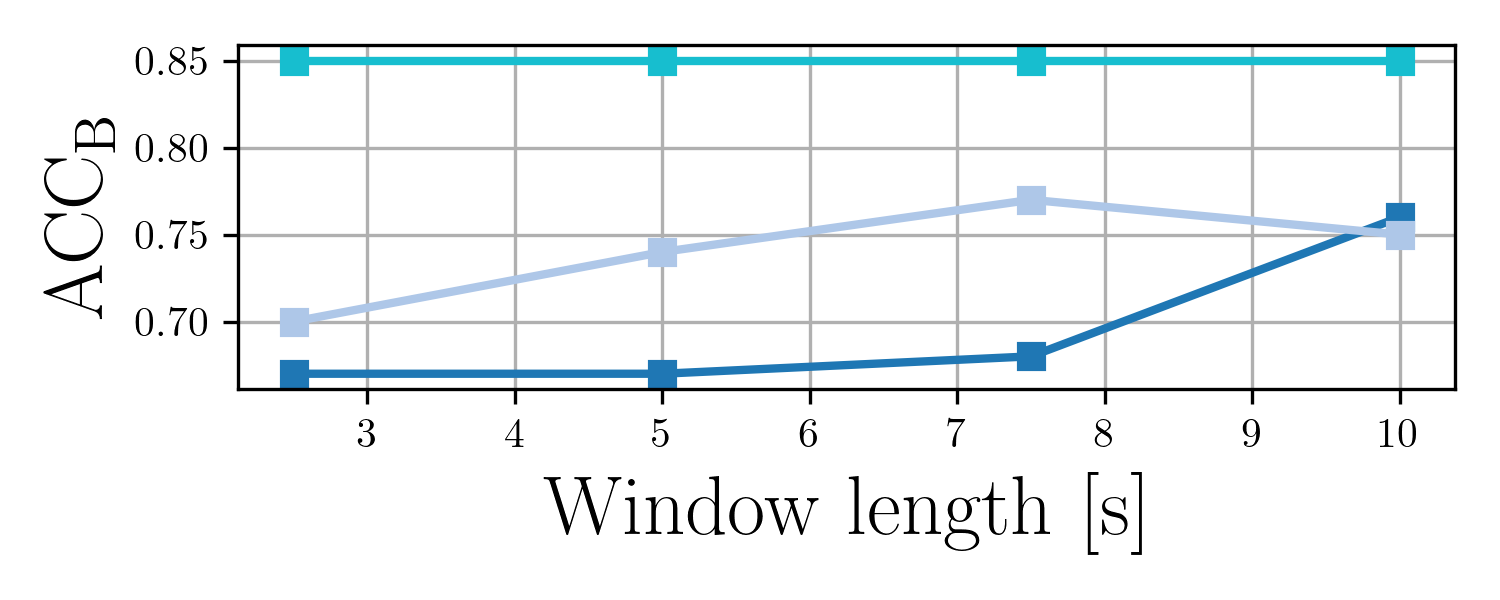}}
\centerline{(b) Open set - Threshold}
\end{minipage}
\hfill
\begin{minipage}[c]{0.32\linewidth}
  \centering
\centerline{\includegraphics[width=\columnwidth]{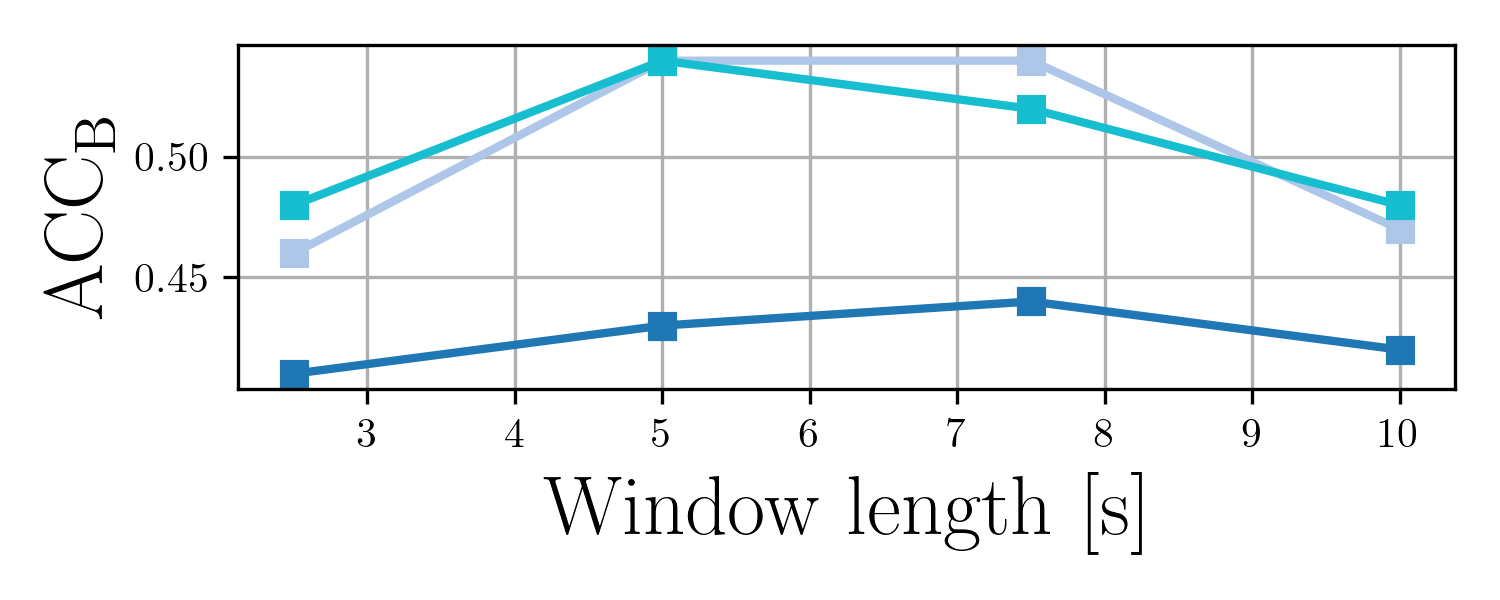}}
\centerline{(C) Open set - SVM}
\end{minipage}

\caption{Balanced accuracy varying according to the considered window size using M5 ({\color{M5} \rule{0.02\linewidth}{1mm}}), RawNet2 ({\color{RAWNET2} \rule{0.02\linewidth}{1mm}}) and ResNet + Spec ({\color{RESNETSPEC} \rule{0.02\linewidth}{1mm}}).}
\label{fig:exp_window_size}
\end{figure*}
We further perform a small experiment to verify how much the impact of the temporal window length used as input to the models changes their performance. This is important, especially for the design of further datasets, i.e. do we need to create longer musical excerpts or not? We consider four window lengths, namely $10~\mathrm{s}$, $7.5~\mathrm{s}$, $5 \mathrm{s}$ and $2.5~\mathrm{s}$ and report the results in terms of balanced accuracy in Fig.~\ref{fig:exp_window_size}. As we can see, the variations in accuracy are not extreme in all classification scenarios. M5 seems to have an accuracy boost passing from $7.5$ to $10 \mathrm{s}$ window length for both closed set and open set (threshold) methods,  ResNet18+Spec does not have major improvements, while a slight increase in accuracy is seen for RawNet2. Results in the case of Open set (SVM) show a less clear trend, but the impact of the window size does not seem to be relevant even in this case.

\subsection{Discussion}
The objective of the results provided in this paper is to present a first approach to the TTM model detection and attribution and do not claim at all to be definitive. Instead, we hope to further motivate research in this direction. New TTM models are proposed almost monthly if not daily, with a continuous increase in quality, especially for what concerns commercial models. For these reasons, while from the results indicated in this paper the problem may seem to be relatively easy, especially in the closed-set scenario, things are not going to stay that way for long and the research community needs to prepare in advance to solve problems related to the detection of fake music.  We can already identify a few developments not analyzed in this paper that could be considered in future works related to TTM attribution. For example, \textit{Do the textual descriptions have an effect on the classification performance?} If text and music are effectively mutually dependent, in the scenario of TTM models, we could be able to leverage on that. Also, \textit{Can we leverage music theory and musicology to detect music deepfakes?} The analysis of musical theory could be of interest in a context where generated music is inserted in otherwise ``real" music, a problem denoted as \textit{splicing.}

\section{Conclusion}
\label{sec:conclusion}
In this paper, we tackled the problem of detecting and attributing music generated via Text-to-music models. Specifically, we introduced the FakeMusicCaps dataset, created by replicating the MusicCaps dataset via five state-of-the-art TTM models. By applying simple audio forensics techniques, we demonstrate that the dataset could be used as an initial benchmark to tackle TTM detection and attribution. Our results are not to be considered definitive, instead, our objective is to further motivate the research in forensics techniques for the analysis of generated music. In fact, while the problem of fake music detection and attribution is now relatively simple, it is guaranteed to grow more extremely complicated day by day.

\bibliographystyle{IEEEtran}
\bibliography{biblio}

\end{document}

%% file: helpers.tex
\usepackage{glossaries}
\newacronym{cnn}{CNN}{Convolutional Neural Network}
\newacronym{dl}{DL}{Deep Learning}
\newacronym{ai}{AI}{Artificial Intelligence}
\newacronym{dnn}{DNN}{Deep Neural Network}
\newacronym{nn}{NN}{Neural Network}
\newacronym{mmf}{MMF}{MultiMedia Forensics}
\newacronym{vc}{VC}{Voice Conversion}
\newacronym{tts}{TTS}{Text-to-Speech}
\newacronym{ann}{ANN}{Artificial Neural Network}
\newacronym{df}{DF}{Deepfake}
\newacronym{ml}{ML}{Machine Learning}
\newacronym{mfcc}{MFCC}{Mel Frequency Cepstral Coefficient}
\newacronym{stft}{STFT}{Short Time Fourier Transform}
\newacronym{cqcc}{CQCC}{Constant Q Cepstral Coefficient}
\newacronym{roc}{ROC}{Receiver Operating Characteristic}
\newacronym{auc}{AUC}{Area Under the Curve}
\newacronym{tp}{TP}{True Positive}
\newacronym{tn}{TN}{True Negative}
\newacronym{fp}{FP}{False Positive}
\newacronym{fn}{FN}{False Negative}
\newacronym{tpr}{TPR}{True Positive Rate}
\newacronym{fpr}{FPR}{False Positive Rate}
\newacronym{tnr}{TNR}{True Negative Rate}
\newacronym{vae}{VAE}{Variational Autoencoders}
\newacronym{gan}{GAN}{Generative Adversarial Networks}
\newacronym{stlt}{STLT}{Short-Term and Long-Term}
\newacronym{gru}{GRU}{Gated recurrent unit}
\newacronym{ser}{SER}{Speech Emotion Recognition}
\newacronym{dft}{DFT}{Discrete Fourier Transform}
\newacronym{mae}{MAE}{Mean Absolute Error}
\newacronym{svm}{SVM}{Support Vector Machines}
\newacronym{gmm}{GMM}{Gaussian Mixture Models}
\newacronym{gnn}{GNN}{Graph Neural Network}
\newacronym{enf}{ENF}{Electric Network Frequency}
\newacronym{nlp}{NLP}{Natural Language Processing}
\newacronym{cm}{CM}{countermeasure}
\newacronym{mlp}{MLP}{Multi-Layer Perceptron}
\newacronym{ssl}{SSL}{Self-supervised Learning}
\newacronym{pf}{PF}{Partial Fake}
\newacronym{fft}{FFT}{Fast Fourier Transform}
\newacronym{eer}{EER}{Equal Error Rate}
\newacronym{ola}{OLA}{Overlap-and-Add}
\newacronym{lfcc}{LFCC}{Linear Frequency Cepstral Coefficient}
\newacronym{fc}{FC}{Fully Connected}
\newacronym{blstm}{BLSTM}{Bidirectional Long Short-Term Memory}
\newacronym{cqt}{CQT}{Constant-Q Transform}
\newacronym{vad}{VAD}{Voice Activity Detector}
\newacronym{add}{ADD}{Audio Deepfake Detection}